\documentclass[prl,aps,twocolumn,showpacs,floatfix,superscriptaddress]{revtex4-1}

\usepackage{graphicx}
\usepackage{dcolumn}
\usepackage{bm}
\usepackage{latexsym}
\usepackage{amsmath}
\usepackage{amssymb}
\usepackage[latin1]{inputenc}
\usepackage[usenames]{color}

\begin{document}

\title{Dielectric spectroscopy of water at low frequencies: The existence of an isopermitive point}

\author{A. Angulo-Sherman and H. Mercado-Uribe}
\email[Corresponding author, ]{Email: hmercado@cinvestav.mx}
\affiliation{CINVESTAV-Monterrey, PIIT, Apodaca, Nuevo Le\'on,  66600 M\'exico}


\begin{abstract}
We have studied the real part of the dielectric constant of water from 100 Hz to 1 MHz.  
We have found that there is a frequency where the dielectric constant is independent of temperature, and 
called this the isopermitive point. Below this point the dielectric constant increases with temperature, above, it decreases. 
To understand this behavior, we consider  water as a system of two species: ions and dipoles. The first give rise to the so called Maxwell-Wagner-Sillars effect, the second obey the Maxwell-Boltzmann statistics. 
At the isopermitive point the effect of both mechanisms in the dielectric response compensate each other.
\end{abstract}

\maketitle

\section{Introduction}

Water, the most abundant substance in Earth, is fundamental for life, and for this reason it has been widely studied for a long time. Due to their atomic and molecular interactions, water is an excellent solvent for many substances \cite{Cabane}. Water molecules not only joint themselves to form tetrahedral arrangements induced by hydrogen bonds, but they can do the same with other molecules. Biological functions are possible only in the presence of water. It is necessary for cell dynamics and plays a relevant role in the stability and flexibility of biomolecules \cite{Chaplin, Oleinikova}. Indeed, many studies have been performed in order to understand the mechanisms of water-protein interaction. In this context, it has been always important to evaluate how the hydration level of proteins influences their conformational motions. In general, water-protein interaction takes place in two tiers: one corresponding to water molecules interacting directly with proteins and the other through weaker forces in bulk water \cite{Pethig, Frauenfelder}. 

One of the most powerful techniques to study the molecular response of water is dielectric spectroscopy. The macroscopic polarizability of a material can be characterized in terms of its relative permittivity \cite{Blythe}. In order to avoid the problem of ionic conduction in water, the dielectric studies are usually  performed at frequencies higher than 1 GHz \cite{Ellison, Kupfer, Lewowski}. It is widely accepted that a reference value of the relative permitivity  is $80$ (at $20 ^{\circ}$C) in the entire region of low frequencies (less than 1 GHz), and only when the temperature changes, the dielectric constant changes \cite{Kupfer, Lide, Katy}. This number is crucial to explain salt solubility in water and the origin of cell functioning.
In the present Letter, we experimentally revisit this subject, measuring the relative permittivity of water at low frequencies. We found that the behavior of this parameter  at low frequencies may have an importance not considered before. Using the dielectric spectroscopy technique we measured the temperature dependence of the dielectric constant in the frequency region from $10^2$ to $10^6$ Hz. Contrary to the expectations, we found that the relative permittivity of water varies as a function of frequency. Moreover, we found that at a well defined frequency its value is independent of temperature. We propose to call this point  the isopermitive point. Below this point the dielectric constant increases with temperature, above, it decreases.  In order to understand this behavior, we consider that  water can be seen as a system of two species: ions and dipoles at different concentrations. So, similarly to the existence of an isosbestic point in reversible chemical processes, where two species exchange concentrations under a temperature change \cite{Robinson}, at the isopermitive point the effect of both dielectric contributions compensate each other.

\section{Experimental}
\subsection{\emph{Setup}}

The experimental setup used in this work is similar to the one employed in a previous study \cite{Gomez}. The central piece of our system is a stainless steel cylindrical capacitor with length $25$ cm consisting of an inside electrode with diameter  $0.47$ cm separated by $0.23$ cm from the outside electrode, whose wall has $0.11$ cm of width. The capacitor is surrounded by cylindrical Nylon shell and all together is put inside a coil made of a cooper pipe.  The whole piece, in turn,  is placed within an aluminum tube and fixed by two taps of the same material at the ends. The electrodes are connected with cooper wires to BNC connectors located at the outside of the taps. In order to change in a controlled way the temperature of the system, water flows through the coil.  The outer aluminum tube not only sets the working temperature but also acts as a Faraday cage to electrically isolate the capacitors and connectors from the outside environment. The temperature of the device (the aluminum tube with the cooper coil and capacitor inside) is controlled using a Thermo Electron heater (Haake DC10), that pumps water continuously. Because the measurements are very sensitive to temperature fluctuations, the aluminum tube was wrapped with a roll of $7$ cm fiberglass and placed inside of a closed cooler. A thermocouple is connected at the wall of the copper pipe and monitored with a multimeter and a computer. The precision in the temperature measurements was about $0.2 ^{\circ}$C. The relative permittivity  $(k^\prime)$ of the water samples is measured by a LCR meter model $4284$A, which operates in the range of $20$ Hz-$1$ MHz, with an accuracy of $\pm  0.05\%$ and $\pm 0.0005$, respectively. The LCR meter has been wrapped in a wire netting to further reduce electromagnetic noise.

\begin{figure}[ht!]
\begin{center}
\includegraphics[scale=0.38]{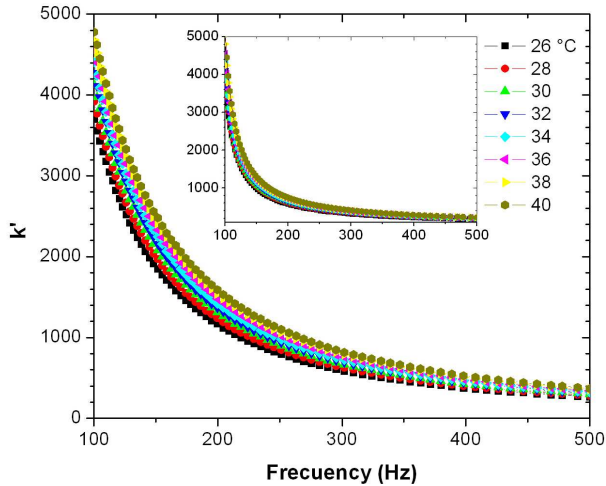}
\caption{\small Relative permittivy of water as a function of frequency at different temperatures: Low frequencies. The inset shows the results obtained using a phenomenological model, discussed in the text. These particular plots represent a single experiment. }
\label{fig1}
\end{center}
\end{figure}

\subsection{\emph{Methods}}

The capacitor is totally filled with $11.9$ ml of bi-distilled water($18.2 M \Omega$cm) in every measurement. The measurements start when the system is at $26 ^{\circ}$C, and thereafter, the temperature increases each $2 ^{\circ}$C until $40 ^{\circ}$C. Every $2 ^{\circ}$C we measure the value of the capacitance. After each temperature increment, we wait approximately $10$ min to reach thermal equilibrium in the whole capacitor. Every  measurement is performed in a frequency range from $100$ Hz - $1$  MHz, with an oscillation voltage of $1$ V. Due to the fact that the capacitor is much longer than the gap between the electrodes, it is not necessary to use a guarded electrode. The relative permittivity is directly the ratio of the measured capacitances with the capacitor filled with water and with the capacitor empty.

\section{Results and discussion}

\begin{figure}[ht!]
\begin{center}
\includegraphics[scale=0.37]{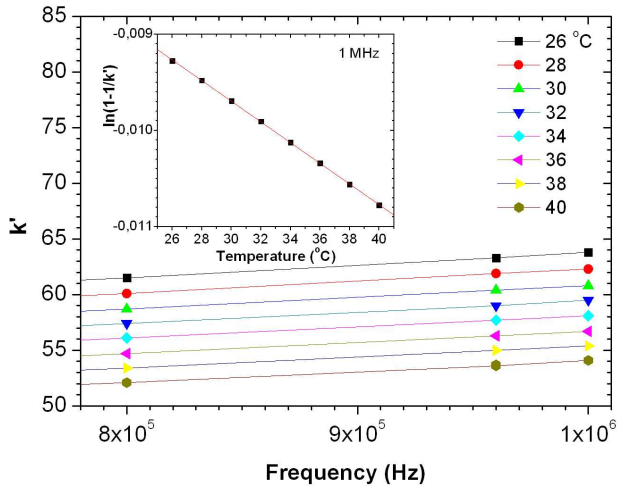}
\caption{\small Relative permittivy of water as a function of frequency at different temperatures: High frequencies. See the text for an explanation of the inset. }
\label{fig2}
\end{center}
\end{figure}

Figs. \ref{fig1} and \ref{fig2} show the frequency dependence of the relative permittivity $(k^\prime)$ of water measured at different temperatures, from $26$ to $40 ^{\circ}$C. We can distinguish two behaviors: On one hand we observe that  $k^\prime$ has an enormous value at $100$ Hz and it reduces as the frequency increases, see Fig. \ref{fig1}. On the other hand, at high frequencies, near $1$ MHz,  $k^\prime$ is almost constant, see Fig. \ref{fig2} . Note that in each regime the temperature has inverse effects: While a temperature increment makes $k^\prime$ to augment in the first regime, it causes a decrease in the second.

Water molecules can auto dissociate forming ion pairs ($H^+$ and $OH^-$) \cite{Trout, Yagasaki, Phillips}. If the temperature is increased, the number of ions from the separation of molecules increases too and a gradual decrease of the pH is obtained \cite{Shiue}. When an oscillating electric field is applied to the  water sample contained in the capacitor, different dielectric processes can occur depending on the frequency of the field. At low frequencies,  a phase difference appears between the electrodes as a result of two dielectric phenomena which ocurr simultaneously, the dipoles orient in the direction of the applied field while the ionic pairs move towards the electrodes (Maxwell-Wagner-Sillars phenomenon \cite{Jansson}). Even if the number of ion pairs is much lower than water molecules, the Maxwell-Wagner-Sillars phenomenon is dominant (this will be discussed later). This is the reason why, under a temperature increase, the relative permittivity increases. At high frequencies, the polarity of the electrodes changes so quickly that only the water dipoles are able to respond effectively, and therefore, the value of the relative permittivity follows the Maxwell-Boltzmann statistics.  As the thermal energy is increased, it becomes more 
and more difficult for the dipoles to align in the direction of the field and the dielectric constant decreases.

\begin{figure}[ht!]
\begin{center}
\includegraphics[scale=0.42]{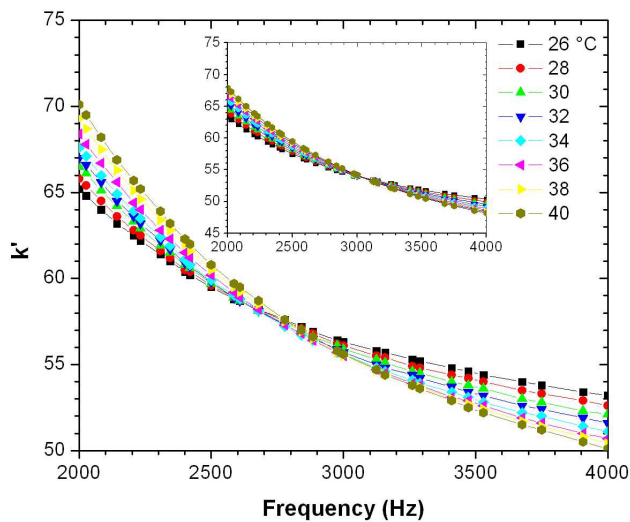}
\caption{\small The existence of an isopermitive point of water: Frequency plots at different temperatures cross in a point. The inset shows the results obtained using a phenomenological model, discussed in the text.} 
\label{fig3}
\end{center}
\end{figure}

In order to model the observed behavior, let us recall that the capacitance we measure is defined by $C  = Q_l/\Delta V$, where $Q_l$ is the  free charge and $\Delta V$ the applied potential difference  between the electrodes. $C$ is usually scaled to the vacuum capacitance $C_0$ (without dielectric medium). The result of this scaling is the dielectric constant $k^\prime$, which is the real part of the complex  permittivity given by $k^{*} =  k^\prime - \imath k^{\prime \prime}$ \cite{Blythe}.
Taking into consideration the polar and ionic sources of polarization,  $k^\prime$ can be written as $\frac{\sigma_l}{\sigma_l-\sigma_p-\sigma_i}$, where $\sigma_l$, $\sigma_p$ and $\sigma_i$ are, respectively, the free, polarized and ionic charge densities. Since $\sigma_p$ and $\sigma_i$ obey different statistics, $k^\prime$ can be represented by the following general expression:

\begin{equation}
k^\prime = \frac{1}{1-ae^{-bT}-F(f,T)}
\end{equation}

The second term in the denominator comes from the Maxwell-Boltzmann distribution, and it is always present in the full range of frequencies. $F(f,T)$, which represents the   small ion contribution, tunes the polarization in the electrodes making $k^\prime$ almost to diverge at low frequencies (i.e. the sum of the second and third term of the denominator approach to one).
$a$ and $b$ can be easily found by fitting $ae^{-bT}$ to $1-\frac{1}{k^\prime}$ at 1 MHz (see the inset of Fig. \ref{fig2}). In such a fit $F(f,T)$ is zero.
We obtained $a = 0.987772$ and $b = 1.05277 \times 10^{-4}/ ^{\circ}C$. $F(f,T)$ can be expressed as $\beta e^{-\gamma f}$ in order to guarantee the reduction of the ionic charge accumulation at high frequencies. Moreover, since the process of water dissociation is a function of temperature (the higher the temperature the lower the pH), $\beta$ and $\gamma$  are functions of  $\frac{1}{T}$. We found that $\beta = 0.021 + 0.034 e^{-78.25/T}$ and $\gamma = 0.0000413 + 0.017/T$. 

The insets of Figs. \ref{fig1} and \ref{fig3} show  the values of the relative permittivity calculated with the above model. We see that the agreement is reasonable. However, further studies will be carried out to find $F(f,T)$ from first principles.

\begin{figure}[ht!]
\begin{center}
\includegraphics[scale=0.35]{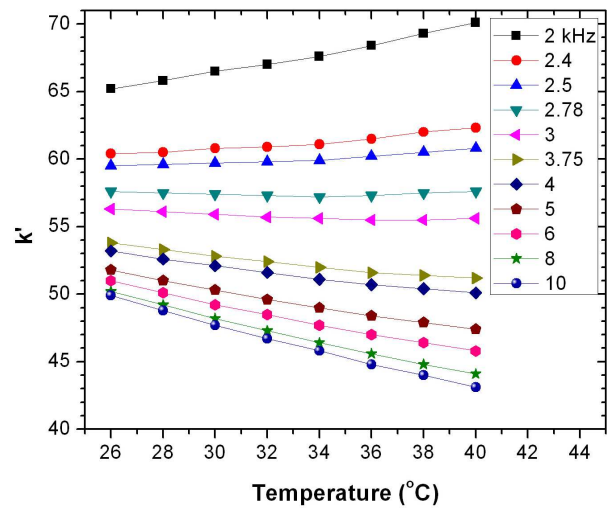}
\caption{\small  $k^\prime$ as a function of temperature around the isopermitive point. The line with the slope equal to zero defines the frequency of the isopermitive point.}
\label{fig4}
\end{center}
\end{figure}

The most striking result of this work is that the observed mechanisms at low and high frequencies give rise to effects that compensate each other as the temperature is varied. In other words, at one point (frequency),  the relative permittivity is independent of temperature, as shown in Fig. \ref{fig3}. In the spectroscopy of reversible chemical reactions a similar phenomenon occurs when two chemical species exchange concentrations under temperature changes and their associated spectra cross at one point. This is called the isosbestic point \cite{Brynestad,Robinson}. Because the term means the same absorbance, it is also widely used in Raman spectroscopy \cite{Walrafen, DArrigo}. However, in such a case, it would be more appropriate to call it, as in reference \cite{Walrafen}, isokedastic (equal scattering) point. Considering these antecedents, we name the crossing point found in our measurements shown in Fig. \ref{fig3}, the isopermitive point. In this sense, water can be seen as a system of two species, ions and dipoles, and in such a point these species contribute independently in such a way that theirs effects compensate each other.

The frequency of the isopermitive point is more precisely located plotting the relative permittivity as a function of temperature for fixed values of frequencies around this point, as  shown in Fig. \ref{fig4}. One can see that the isopermitive point is  where the slope is practically zero. We repeated these measurements with differents water samples and found that this value is at $2.96\pm 0.17$ kHz.
We would like to emphasize that although the value of the relative permittivity of water below $10^9$ Hz is considered to be unchanged as a function of frequency \cite{Lide}, we have found that it is important to measure $k^\prime$ at low frequencies. 
We believe that the existence of the so called isopermitive point could have implications in understanding biological processes because a biological cell can be seen as a capacitor where there is an exchange of ions through the membrane. We would like to remark that during cell functioning the transport of ions occur at a frequency of the order of $1$ kHz. 

\section{Conclusions}
We have  measured the dielectric constant of water between $100$ Hz to $1$ MHz at different temperatures. We have found that there is a frequency where the value of the dielectric constant is independent of temperature, and called this the isopermitive point.  This point adds to other physical crossing points insensitive to temperature like the isosbestic point found in chemical reactions or Raman spectroscopy. Furthermore, the frequency of this intruiging point is very close to the frequency of a recently discovered relaxation process in water  \cite{Jansson}.

\begin{center}
Acknowledgments
\end{center}
The authors would like to thank C. Ruiz for the valuable comments and proofreading of the manuscript and J. Salda\~na for his help in building the capacitor. A.A.S. wishes to acknowledge a scholarship from Conacyt, Mexico.


\end{document}